# Optimal Entanglement Provision over Underlying Networks of Group Association Schemes

M. Ghojavand [1], M. A. Jafarizadeh [2, 3] and S. Rouhani [1]

1 Physics Department, Sharif University of Technology, P.O. Box 11365-9161, Tehran, Iran
2 Department of Theoretical Physics and Astrophysics, University of Tabriz, Tabriz 51664, Iran
3 Research Institute for Fundamental Sciences, Tabriz 51664, Iran

**ghojavand@gmail.com**, **jafarizadeh@tabrizu.ac.ir**, **srouhani@sharif.ir**

**Abstract**

We investigate creation of optimal entanglement in symmetric networks. The proposed network is one in which the connections are organized by a group table in other words a group association schemes. We show that optimal entanglement can be provided over such networks by adjusting couplings and initial state. We find that the optimal initial state is a simple local excited state. Moreover we obtain an analytic formula for entanglement achievable between vertex pairs located in the same strata and calculate their relevant optimal couplings. For those pairs located in different strata two upper bounds are presented. These results are also checked numerically.

## 1   Introduction

Coupled spin networks have received considerable attention for applications in low dimensional quantum information processors. In these systems fundamental actions such as high fidelity transfer of quantum state or entanglement provision [1][2][3][4]could occur through natural evolution of the initial state of the system without any ancillary external manipulation [5][6][7][8]. Recently graph theory has contributed to this area, for example perfect state transfer (PST) between antipodal vertices, in some symmetric graphs has been demonstrated [9][10][11]. A class of graphs that demonstrate this capability are Underlying Networks of Group Association Schemes (UNGAS) [11]. These networks are organized such that each vertex pair is connected according to the group multiplication table, on each vertex sits a spin variable. Thus one has a coupled spin network, with connections organized according to a group multiplication table. On the other hand symmetric graphs are best suited for entanglement provision between different vertex pairs [6]. Due to intrinsic symmetry of UNGAS we may derive concrete results. We find that the initial optimal state is a single local excited state, the exact amount of optimal entanglement and its related optimal couplings are calculated for some vertex pairs. For the other pairs we can't drive the exact amount but we present two upper bounds for their optimal amount of entanglement, exact expressions may be found below in Eqs. (4-4), (4-6), (4-13), (4-15), (5-4) and (5-6).

The organization of this paper is as follows: in section 2 we recall some preliminary material on association schemes and their underlying networks, and in particular UNGAS. In section 3 we consider the case of qubits interacting via the map of UNGAS and compute the entanglement generation between a given pair of vertices. In sections 4 and 5 we consider conditions for optimal entanglement generation between vertex pairs. In section 4 we limit ourselves to pairs with equal excitation amplitude, whereas in section 5 we lift this condition, some examples are given. The paper is closed with a summary in section 6.

## 2    Underlying networks of Group association schemes (UNGAS)

### 2.1    Association schemes

Association schemes were originally introduced by Bose and his co-workers in the design of statistical experiments[13][14]. Since their inception, the concept has proved useful in many areas [12] such as the study of group actions in algebraic graph theory, algebraic coding theory [15], knot theory [16] and numerical integration [17]. In the following, we present some preliminary material on Underlying Networks of Association Schemes (UNAS) and Underlying Networks of Group Association Schemes (UNGAS); for more detailed information about association schemes and their underlying networks, refer to [12][18][19][20].

**Definition 1** let $V$ be a set of n vertices and assume $\{R_i : i = 0,1,...,d\}$ be a set of non empty relations on $V$ (subsets of $V \times V$) satisfying the following conditions (AS1) to (AS4). Then the pair $Y = \{V, R_i : 0 \leq i \leq d\}$ is called an association scheme. The constraints on relations are

(AS1)  $\{R_i : 0 \leq i \leq d\}$ is a partition of $V \times V$

(AS2)  $R_0 = \{(\alpha, \alpha) : \alpha \in V\}$

(AS3)  $R_i^t \in \{R_j : 0 \leq j \leq d\}$ , where $R_i^t := \{(\beta, \alpha) : (\alpha, \beta) \in R_i\}$

(AS4)  For $(\alpha, \beta) \in R_k$, the number $p_{i,j}^k = |\{\gamma \in V : (\alpha, \gamma) \in R_i \ \& \ (\gamma, \beta) \in R_j\}|$ does not depend on $(\alpha, \beta)$ but only on $i, j$ and $k$.

If

(AS3)´  $R_i^t = R_i$

then (AS3) is satisfied and the corresponding association scheme is called symmetric. In symmetric association schemes condition (AS3)´ implies that

$$p_{i,j}^k = p_{j,i}^k \tag{2-1}$$

Considering constraint (AS4) it can be shown that the set of adjacency matrices defined as

$$(\mathbf{A}_i)_{\alpha,\beta} = \begin{cases} 1 & \text{if } (\alpha, \beta) \in R_i \\ 0 & \text{otherwise} \end{cases} \tag{2-2}$$

obey the Bose-Mesner algebra which is:

$$\mathbf{A}_i \mathbf{A}_j = \sum_{k=0}^{d} p_{i,j}^k \mathbf{A}_k \tag{2-3}$$

In symmetric association schemes, according to (2-1) and (2-3), the set of these matrices are commutative and regarding their hermiticity they can simultaneously be diagonalized in a common orthonormal basis. So the adjacency matrices $\mathbf{A_0}, \mathbf{A_1}, \ldots, \mathbf{A_d}$ may be expanded by a new set of projection operators (idempotents) $\mathbf{E_0}, \mathbf{E_1}, \ldots, \mathbf{E_d}$ where $\mathbf{E_i}$ project the state to $i$-th common eigenvector, so

$$\mathbf{E}_0 = \frac{1}{n}\mathbf{J}, \quad \mathbf{E}_i \mathbf{E}_j = \delta_{ij} \mathbf{E}_i, \quad \sum_{i=0}^{d} \mathbf{E}_i = \mathbf{1} \tag{2-4}$$

(here $J$ is an $n \times n$ matrix where all of its elements are 1)

Let $P$ and $Q$ be two $(d+1) \times (d+1)$ matrices which are the expansion coefficients of $\{A_i\}$ by $\{E_i\}$ and vice versa so:

$$\mathbf{A}_j = \sum_{i=0}^{d} P_{ij} \mathbf{E}_i \tag{2-5}$$

$$\mathbf{E}_i = \sum_{i=0}^{d} Q_{ji} \mathbf{A}_j \tag{2-6}$$

Here $P_{ij}$ (resp. $Q_{ij}$) is the $j$-th eigenvalue (respectively the $j$-th dual eigenvalue) of $\mathbf{A}_i$ (resp. $\mathbf{E}_i$) and that the columns of $\mathbf{E}_j$ are the corresponding eigenvectors. Clearly we have

$$PQ = QP = nI \tag{2-7}$$

Clearly, each non-diagonal relation $R_i$ of a symmetric association scheme can be thought of as undirected networks $\Gamma_i = (V, R_i)$ on $V$, which we consider as the underlying networks which describe the interactions between different nodes. The relationship between UNGAS and a physical system is that different vertex pairs that belong to the same relation $\Gamma_i$ have equal coupling strength $J_i$ (i.e. 3-1).

## 2.2 Network stratification and its Hilbert space

Consider $\Gamma_i(o) = \{\beta \in V : (o, \beta) \in R_i\}$ for an arbitrary chosen vertex $o \in V$ (called reference vertex), then the vertex set V can be written as disjoint union of $\Gamma_i(o)$, i.e.

$$V = \bigcup_{i=0}^{d} \Gamma_i(o) \tag{2-8}$$

In fact relation (2-8) stratifies the network into the disjoint union of strata $\{\Gamma_i(o)\}$ (associate classes) each containing $\kappa_i = |\Gamma_i(o)|$ vertices. Now let us define the Hilbert space corresponding to our quantum walk dynamics. Consider the set of $n$ orthonormal kets $\{|i,l\rangle \ni 0 \leq i \leq d; 1 \leq l \leq \kappa_i\}$, where $|i,l\rangle$ means the ket corresponds to the $l$-th vertex of $i$-th stratum. We define the Hilbert space as: $W = span\{|i,l\rangle \ni 0 \leq i \leq d; 1 \leq l \leq \kappa_i\}$. We show that the dynamics of quantum state is confined to a sub-space of $W$. The set of eigenvectors $\{|\phi_i\rangle\}$ defined by

$$|\phi_i\rangle = \frac{1}{\sqrt{\kappa_i}} \sum_{l=1}^{\kappa_i} |i,l\rangle \tag{2-9}$$

spans this subspace. Recall the definition of the adjacency matrices $A_i$ (2-2), it is easy to show that their action on the reference state $|\phi_0\rangle$ is :

$$\mathbf{A}_i |\phi_0\rangle = \sum_{l=1}^{\kappa_i} |i,l\rangle \tag{2-10}$$

Note that $|\phi_0\rangle = |o,1\rangle$ and recall $\kappa_0 = 1$, so for the sake of simplicity hereafter we use the notation $|o\rangle := |o,1\rangle$, So by equation (2-9) we have:

$$\mathbf{A}_i |\phi_0\rangle = \sqrt{\kappa_i} |\phi_i\rangle \tag{2-11}$$

### 2.3 Group association schemes

In order to study quantum walk dynamics on group graphs, we need to study the group association schemes. Let $G$ be a group, then it could be proved that the set of relations defined by

$$R_i = \{(\alpha, \beta) : \alpha\beta^{-1} \in C_i\}, \tag{2-12}$$

where $\{C_i : 0 \leq i \leq d\}$ are the set of conjugacy classes of $G$, satisfying the (AS1)-(AS4) and (AS3)′ conditions, so $Y = \{G; R_i : 0 \leq i \leq d\}$ is a symmetric association scheme.

It is easy to show that $i$-th adjacency matrix is a summation over $i$-th stratum group element. In fact by the action of $\overline{C_i} := \sum_{g \in C_i} g$ ($\overline{C_i}$ is called $i$-th *class sum*) on group elements in regular representation we observe that $\forall \alpha, \beta : (\overline{C_i})_{\alpha\beta} = (\mathbf{A}_i)_{\alpha\beta}$, so

$$\mathbf{A}_i = \overline{C_i} = \sum_{g \in C_i} g \quad . \tag{2-13}$$

Thus due to (2-3)

$$\overline{C_i C_j} = \sum_{k=0}^{d} p_{i,j}^k \overline{C_k} \tag{2-14}$$

However from group theory we know that the coefficients $p_{i,j}^k$ have to be non negative integer, and their value is given by the following relation[21].

$$p_{i,j}^k = \frac{n}{\kappa_i \kappa_j} \sum_{m=0}^{d} \frac{\chi_m(g_i)\chi_m(g_j)}{d_m} \overline{\chi_m(g_k)} \tag{2-15}$$

$$P_{i,j} = \frac{\kappa_j}{d_i} \overline{\chi_i(g_j)} \tag{2-16}$$

$$Q_{i,j} = d_j \overline{\chi_j(g_i)} \tag{2-17}$$

where $d_j := \chi_j(1)$'s are the dimensions of irreducible characters $\chi_j$ which have positive integer values and $\kappa_i := |C_i|$ depicts the number of vertices lying in the $i$-th stratum and $n := |G|$ is the total number of vertices.

## 3  Optimal entanglement between different pairs of qubits interacting via UNGAS

Let us consider a practical situation where $n$ qubits are interacting with each other via the map of an underlying network of association schemes (UNAS). We ask: what is the maximum amount of entanglement which can be generated between two targeted nodes. We use a simple protocol first the reference qubit (reference vertex $o$) is initialized optimally. Then the system is allowed to evolve under the above natural interaction for an optimal time. We look for the optimal interaction coupling in order to increase the entanglement of the targeted pair as much as possible. We assume that the dynamics of the system is governed by the well known ferromagnetic Heisenberg Hamiltonian [22][23] :

$$\mathbf{H} = \sum_{l=0}^{d} J_l \sum_{(i,j) \in R_l} \vec{\sigma}_i \cdot \vec{\sigma}_j \tag{3-1}$$

In the ground state all of the spins are parallel and downward, so $|\underline{0}\rangle = \bigotimes_{j=0}^{n} |0\rangle_i = |00...0\rangle$ where by "0" we mean spin down, and by "1" we mean spin up. Presumably this state is easy to prepare by extreme cooling. At the next step we initialize the quantum state of the reference qubit $o$ as $|\psi\rangle_o = \alpha |0\rangle_o + \beta |1\rangle_o$. The state of the entire spin system after this step is given by

$$|\Psi(t=0)\rangle = (\alpha |0\rangle_o + \beta |1\rangle_o) \bigotimes_{j \neq o}^{n} |0\rangle_i = Sin(\theta)e^{i\varphi}|\underline{0}\rangle + Cos(\theta)|o\rangle \tag{3-2}$$

here we have introduced new real variables $\theta$ and $\varphi$ instead of complex variables $\alpha$ and $\beta$.

Recall that $|o\rangle := |1_o 000...0\rangle$, i.e. all of the spins are down except $o$-th spin which is up.

Since $[\sigma_{Z\_total}, \mathbf{H}] = 0$ the total excitation of spins is conserved in different times. As a matter of fact $\mathbf{H}$ is block diagonal:

$$\mathbf{H} = \mathbf{H}^{(0)} \oplus \mathbf{H}^{(1)} \oplus \mathbf{H}^{(2)} \oplus ... \oplus \mathbf{H}^{(n)} \tag{3-3}$$

However in our case only the two first blocks contribute, so the union of $W$ (the Hilbert space in the last section) and ket $|\underline{0}\rangle$ is sufficient to study the dynamics of this quantum state. Accordingly the quantum state at time t will be

$$|\Psi(t)\rangle = Sin(\theta)e^{i\varphi}|\underline{0}\rangle + Cos(\theta)\sum_{i=0}^{d}\sum_{l=1}^{\kappa_i} f_{il}(t)|i,l\rangle \tag{3-4}$$

where by the definition $f_{il}(t) := \langle i,l | e^{\frac{-i\mathbf{H}t}{\hbar}} | o \rangle$ is the transition amplitude. Recall that $f_{il}(t)$ expresses the transition probability amplitude of the state $|o\rangle$ at $t_0 = 0$ to the state $|i,l\rangle$ at time t. For the sake of simplicity from here on we set $\hbar = 1$. Due to probability conservation we have

$$\sum_{i=0}^{d}\sum_{l=1}^{\kappa_i} |f_{il}(t)|^2 = 1 \tag{3-5}$$

In order to compute the entanglement generated between two vertices $(i,l)$ and $(i',l')$ we must determine the reduced density matrix of these two qubits. Tracing over all vertices except $(i,l)$ and $(i',l')$ we get:

$$\boldsymbol{\rho}_{il,il'}(t) = \begin{pmatrix} 1-Cos^2(\theta)(|f|^2+|f'|^2) & \frac{1}{2}Sin(2\theta)e^{-i\varphi}f' & \frac{1}{2}Sin(2\theta)e^{-i\varphi}f & 0 \\ \frac{1}{2}Sin(2\theta)e^{i\varphi}f'^* & Cos^2(\theta)|f'|^2 & Cos^2(\theta)ff'^* & 0 \\ \frac{1}{2}Sin(2\theta)e^{i\varphi}f^* & Cos^2(\theta)f'f^* & Cos^2(\theta)|f|^2 & 0 \\ 0 & 0 & 0 & 0 \end{pmatrix} \tag{3-6}$$

Here we use the concise notation $f := f_{il}(t)$ and $f' := f_{il'}(t)$. In order to quantify entanglement between two qubits $(i,l)$ and $(i',l')$ we use concurrence of $\boldsymbol{\rho}_{il,il'}(t)$, a measure which quantifies entanglement between two qubits [24] [25]. The concurrence of $\rho$ is defined as $C = \max\{0, \lambda_1 - \lambda_2 - \lambda_3 - \lambda_4\}$, where $\lambda_i$ 's are in decreasing order the square root of eigenvalues of the non-Hermitian matrix $\mathbf{R} = \rho(\sigma_y \otimes \sigma_y)\rho^*(\sigma_y \otimes \sigma_y)$ and $\sigma_y$ is the Pauli spin-flip operator. After some algebraic manipulation the result is

$$C_{il,i'l'}(t) = 2Cos(\theta)|f_{il}(t)||f_{i'l'}(t)| \tag{3-7}$$

Since the transition amplitudes are independent of the initial parameters $\theta$ and $\varphi$ the first step easily reads $\theta_{opt} = 0$. In other words the optimal initial state is

$$|\Psi_{opt}(t=0)\rangle = |o\rangle \tag{3-8}$$

This is good news from experimental point of view because initialization is simply done by applying a local magnetic field in the Z direction at reference site during the cooling process. Moreover only $\mathbf{H}^{(1)}$ in (3-3) governs the dynamics of optimal initial state (3-8).

So, hereafter we shall use (3-8) for optimization of the parameters. Henceforth, our task is to optimize the multiplication of each pair of transition amplitudes

$$T(t) = 2|f_{il}(t)||f_{i'l'}(t)| \tag{3-9}$$

as target function by setting the corresponding optimal couplings in (3-1). This is done where n qubits interacting via the map of an underlying network of a group association scheme (UNGAS).

It is not difficult to show that the governing Hamiltonian $\mathbf{H}^{(1)}$ could be expanded linearly in terms of the set of adjacency matrices (2-2) as follows [9] :

$$\mathbf{H} \equiv \mathbf{H}^{(1)} = 2\sum_{l=0}^{d} J_l \mathbf{A}_l + \frac{n-4}{2} \sum_{l=0}^{d} \kappa_l J_l \mathbf{I} \tag{3-10}$$

Note that the second term in (3-4) is the identity matrix, so it has no effect on the quantum dynamics, therefore it may be ignored.

Let us now compute the transition amplitudes $f_{il}(t)$ in terms of J's for the case of $n$ qubits interacting via UNGAS. Substituting the set of $\mathbf{A}$'s in terms of $\mathbf{E}$'s in (3-10); the unitary evolution operator will be:

$$\mathbf{U}(t) = e^{-i\mathbf{H}t} = e^{-2it\sum_{k=0}^{d} J_k \mathbf{A}_k} = e^{-2it\sum_{k=0}^{d} J_k \sum_{l=0}^{d} P_{lk} \mathbf{E}_l} = \sum_{l=0}^{d} e^{-2it\sum_{k=0}^{d} J_k P_{lk}} \mathbf{E}_l \tag{3-11}$$

where the equality $\left(\sum_{l=0}^{d}\mu_l \mathbf{E}_l\right)^m = \sum_{l=0}^{d}\mu_l^m \mathbf{E}_l$ is used, that can be derived from (2-4). Noting the simple action of adjacency matrices on the initial state (3-8) we substitute $\mathbf{E}_i$'s in terms of $\mathbf{A}_i$'s in (3-11) and get the final form of the quantum state :

$$|\Psi(t)\rangle := e^{-i\mathbf{H}t}|o\rangle = \sum_{l=0}^{d}\frac{1}{n}\sum_{m=0}^{d} e^{-2it\sum_{k=0}^{d}J_k P_{lk}} Q_{ml}\mathbf{A}_m|o\rangle = \sum_{m=0}^{d}\sqrt{\kappa_m}\alpha_m(t)|\phi_m\rangle \qquad (3\text{-}12)$$

Where by definition

$$\alpha_m(t) := \frac{1}{n}\sum_{l=0}^{d} e^{-2it\sum_{k=0}^{d}J_k P_{lk}} Q_{ml} \qquad (3\text{-}13)$$

which is the transition amplitude:

$$f_{jl}(t) := \langle j,l|e^{-iHt}|o\rangle = \alpha_j(t) \qquad (3\text{-}14)$$

Note that the value of transition amplitude $f_{jl}(t)$ depends only on the index $j$ and not on $l$. In other words, all of the vertices located in the same stratum have equal transition amplitudes. Thus the form of target function of entanglement reduces to :

$$T(t) = 2|\alpha_j(t)||\alpha_{j'}(t)| \qquad (3\text{-}15)$$

This means if targeted vertices for entanglement optimization are locate in the same stratum, then the problem of optimization of entanglement becomes equivalent to that of $|\alpha_j(t)|$ (state transfer over $j$-th stratum), which can be treated analytically. So for such vertex pairs we have:

$$C_{ml,ml'}^{opt} = 2|\alpha_m(t)|_{opt}^2 \qquad (3\text{-}16)$$

However if the targeted vertices are located on different strata then the situation is more complex and all we can do is to offer some upper bounds. These bounds may be checked numerically.

## 4 Optimization of state transfer over strata in UNGAS
As we pointed out above, when vertex pairs are located on the same stratum, all we need do is to optimize $|\alpha_j(t)|$. Substituting from (2-16) and (2-17) in (3-13) we get:

$$\alpha_m(t) = \frac{1}{n}\sum_{l=0}^{d} e^{\frac{-2it}{\chi_l(1)}\sum_{k=0}^{d} J_k \kappa_k \overline{\chi_l(g_k)}} d_l \overline{\chi_l(g_m)}$$

$$:= \frac{1}{n}\sum_{l=0}^{d} e^{i(\theta_l - 2\xi_{ml})} d_l |\chi_l(g_m)| \qquad (4\text{-}1)$$

Where $\theta_l$ are defined as

$$\theta_l(t) := -2t \sum_{k=0}^{d} J_k P_{kl} = \frac{-2t}{d_l}\sum_{k=0}^{d} J_k \kappa_k \overline{\chi_l(g_k)} \qquad (4\text{-}2)$$

and $2\xi_{ml}$ is the imaginary phase of $\chi_l(g_m)$.

In equation (4-1) only the phases $\{\theta_l(t)\}$ depend on the optimization parameters $\{J_l\}$. Therefore assuming that all $\theta_l(t)$'s are real, then the optimal value of $|\alpha_m(t)|$ is obtained by adjusting the set of optimization parameters $\{J_l : 0 \leq l \leq d\}$ such that the following equation holds::

$$\theta_l - 2\xi_{ml} = -2t\sum_{k=0}^{d} J_k P_{kl} - 2\xi_{ml} = 2n_l\pi + 2\Phi \quad \text{where } 0 \leq l \leq d \qquad (4\text{-}3)$$

Thus the optimal value of $|\alpha_m(t)|$ reads

$$|\alpha_m(t)|_{opt} = \frac{1}{n}\sum_{l=0}^{d} d_l |\chi_l(g_m)| \qquad (4\text{-}4)$$

Here the set of $\{n_l\}$ are arbitrary integers, and $\Phi$ is a constant real.

The set of equations (4-3) could be written in vector form as

$$(PJ)_l = -\frac{n_{ml}\pi + \xi_{ml} + \Phi}{t_m^*} \qquad (4\text{-}5)$$

Multiplying by matrix $Q$ from left the optimal couplings are obtained as:

$$J_k^{opt} = \frac{-d_k}{nt_m^*}\sum_{l=0}^{d}(n_{ml}\pi + \xi_{ml} + \Phi)\overline{\chi_k(g_l)} \qquad (4\text{-}6)$$

## 4.1 Real and Complex valued character tables in UNGAS

As pointed out, the above strategy for optimization is valid as long as the reality of $\theta_l$'s are assumed. This assumption is valid if the character table is entirely real. Real valued couplings, make physical sense, but characters need not always be real.

Another way to warrant the reality of $\theta_l$'s is to set equal each pairs of couplings corresponding to conjugate columns of the character table:

$$J_l = J_{\bar{l}} \quad \text{if} \quad \chi_k(g_l) = \overline{\chi_k(g_{\bar{l}})} \tag{4-7}$$

Here $\bar{l}$ is the index of column that is complex conjugate of column with index $l$. Recall that the dimension does not change:

$$d_l = d_{\bar{l}} \tag{4-8}$$

Note that the constraints (4-7) get added to the $d+1$ equations already imposed by (4-3), therefore in groups with complex characters the number of independent optimization parameters reduces from $d+1$ to $d' = d+1-I/2$ where $I$ stands for the number of complex valued columns (rows) of complex valued group character table. This means that the equations are over determined and may not admit a solution. Also the solution obtained in (4-4) is no longer valid for groups with complex characters. Let us therefore attempt at finding a new solution. Using the constraint (4-7) and (4-8) in (4-1) we get:

$$\alpha_m(t) := \frac{1}{n} \sum_{l=0}^{R} d_l \chi_l(g_m) e^{\frac{-2it}{d_l}\left(\sum_{k=0}^{R} J_k \kappa_k \chi_l(g_k) + \sum_{k=R+1}^{d'} J_k \kappa_k \left(\chi_l(g_k) + \chi_l(g_{\bar{k}})\right)\right)}$$

$$+ \frac{1}{n} \sum_{l=R+1}^{d'} d_l \left(\chi_l(g_m) + \chi_l(g_{\bar{m}})\right) e^{\frac{-2it}{d_l}\left(\sum_{k=0}^{R} J_k \kappa_k \chi_l(g_k) + \sum_{k=R+1}^{d'} J_k \kappa_k \left(\chi_l(g_k) + \chi_l(g_{\bar{k}})\right)\right)} \tag{4-9}$$

Here we represent the real valued columns (rows) and complex valued columns (rows) distinct from each other.

Now we introduce a new variable:

$$\zeta_l(m) := \begin{cases} \chi_l(g_m) & \text{if } \forall l: \chi_l(g_m) = \overline{\chi_l(g_m)} \\ \chi_l(g_m) + \chi_l(g_{\bar{m}}) & \text{else} \end{cases} \tag{4-10}$$

Where $0 \leq m \leq d'$ and $0 \leq l \leq d'$. Therefore due to (4-7) we have a new $(d'+1) \times (d'+1)$ real valued table $\zeta_l(m)$. This new table is constructed at the first step by merging each conjugate column of character table in its dual column; then proceeds by eliminating the dual row of each complex row in the character table. Using these new real variables in (4-9) we get:

$$\alpha_m(t) := \frac{1}{n} \sum_{l=0}^{d'} d_l \zeta_l(m) e^{\frac{-2it}{d_l} \sum_{k=0}^{d'} J_k \kappa_k \zeta_l(k)} \tag{4-11}$$

The form of (4-11) is similar to (4-1) so for its optimization we can follow exactly the same procedure. Replacing $\chi_l(g_m)$ by $\zeta_l(m)$ in (4-4) and (4-6) we get:

$$|\alpha_m(t)|_{opt} = \frac{1}{n}\sum_{l=0}^{d'} d_l |\zeta_l(m)| \tag{4-12}$$

$$|\alpha_m(t)|_{opt} = \frac{1}{n}\sum_{l=0}^{R} d_l |\chi_l(g_m)| + \frac{1}{n}\sum_{l=R+1}^{d'} d_l |\chi_l(g_m) + \overline{\chi_l(g_m)}| \tag{4-13}$$

$$J_k^{opt} = \frac{-d_k}{nt_m^*}\sum_{l=0}^{d'}(n_{ml}\pi + \xi_{ml} + \Phi)\zeta_k(l) \tag{4-14}$$

$$J_k^{opt} = \begin{cases} \dfrac{-d_k}{nt_m^*}\sum_{l=0}^{d'}(n_{ml}\pi+\xi_{ml}+\Phi)\chi_k(g_l) & 0\le k \le R \\ \dfrac{-d_k}{nt_m^*}\sum_{l=0}^{d'}(n_{ml}\pi+\xi_{ml}+\Phi)(\chi_k(g_l)+\overline{\chi_k(g_l)}) & R < k \le d' \end{cases} \tag{4-15}$$

We observe that due to (4-9), $\alpha_m(t)$ are invariant under the transformation $m$ to $\overline{m}$ since $\chi_l(g_m) = \overline{\chi_l(g_{\overline{m}})}$ so we have

$$\alpha_m(t) = \alpha_{\overline{m}}(t) \tag{4-16}$$

Consequently:

$$\{J_l\}_{opt}^m = \{J_l\}_{opt}^{\overline{m}} \tag{4-17}$$

Equation (4-17) means that if the optimal value of excitation amplitude over stratum $m$ is achieved optimization over its dual stratum $\overline{m}$ has happened as well. In the next subsection we give some examples and calculate the optimal value of excitation amplitudes over different strata.

### 4.2 Examples of optimal state transfer over strata in UNGAS

In this section we investigate optimal entanglement between vertices located in the same strata in some easy to handle groups.

### 4.2.1 The Dihedral group $D_6$

The Dihedral group $G = D_{2s}$ [21] is generated by two generators $a$ and $b$ with the following relations: [21]

$$D_{2s} = \{a,b : a^s = b^2 = 1, b^{-1}ab = a^{-1}\}$$

The number of group elements of $D_6$ is $n = 6$ and the character table of this group is as follow:

| $g_i$ | 1 | $a$ | $b$ |
|---|---|---|---|
| $\kappa_i$ | 1 | 2 | 3 |
| $\chi_1$ | 1 | 1 | 1 |
| $\chi_2$ | 1 | 1 | −1 |
| $\chi_3$ | 2 | −1 | 0 |

Due to reality of character table of $D_6$ by using (4-4) and (2-16) the optimal value of the $|\alpha_i|_{opt}$ obtained as:

$$|\alpha_0|_{opt} = \frac{1}{6}(1\times1+1\times1+2\times2) = 1$$

$$|\alpha_1|_{opt} = \frac{1}{6}(1\times1+1\times1+2\times1) = \frac{2}{3}$$

$$|\alpha_2|_{opt} = \frac{1}{6}(1\times1+1\times1+2\times0) = \frac{1}{3}$$

Using (3-16) we get

$$C^{opt}_{1l,1l'} = \frac{8}{9}$$

$$C^{opt}_{2l,2l'} = \frac{2}{9}$$

### 4.2.2 The Cyclic group $Z_{2k}$

The even cyclic group $G = Z_{2k}$ is generated by one generator $a$ with the following relations [21]

$$Z_{2k} = \{a : a^{2k} = 1\}$$

The number of group elements of $Z_{2k}$ is 2k and the character table of this group is as follow:

| $g_i$ | $e$ | $a$ | $a^2$ | . . . | $a^j$ | . . . | $a^{k-1}$ | $a^k$ |
|---|---|---|---|---|---|---|---|---|
| $\kappa_i$ | 1 | 2 | 2 | . . . | 2 | . . . | 2 | 1 |
| $\chi_1$ | 1 | 1 | 1 | . . . | 1 | . . . | 1 | 1 |
| $\chi_2$ | 1 | $-1$ | 1 | . . . | $(-1)^j$ | . . . | $(-1)^{k-1}$ | $(-1)^k$ |
| $\chi_3$ | 1 | $\omega$ | $\omega^2$ | . . . | $\omega^j$ | . . . | $\omega^{k-1}$ | $(-1)^k$ |
| $\chi_4$ | 1 | $\bar{\omega}$ | $\bar{\omega}^2$ | . . . | $\bar{\omega}^j$ | . . . | $\bar{\omega}^{k-1}$ | $(-1)^k$ |
| . | . | . | . | | . | | . | . |
| . | . | . | . | | . | | . | . |
| $\chi_{2l-1}$ | 1 | $\omega^l$ | $\omega^{2l}$ | . . . | $\omega^{jl}$ | . . . | $\omega^{(k-1)l}$ | $(-1)^{kl}$ |
| $\chi_{2l}$ | 1 | $\bar{\omega}^l$ | $\bar{\omega}^{2l}$ | . . . | $\bar{\omega}^{jl}$ | . . . | $\bar{\omega}^{(k-1)l}$ | $(-1)^{kl}$ |
| . | . | . | . | | . | | . | . |
| . | . | . | . | | . | | . | . |
| $\chi_{2k-1}$ | 1 | $\omega^k$ | $\omega^{2k}$ | . . . | $\omega^{jk}$ | . . . | $\omega^{(k-1)k}$ | $(-1)^{k^2}$ |
| $\chi_{2k}$ | 1 | $\bar{\omega}^k$ | $\bar{\omega}^{2k}$ | . . . | $\bar{\omega}^{jk}$ | . . . | $\bar{\omega}^{(k-1)k}$ | $(-1)^{k^2}$ |

where $\omega := e^{\frac{\pi i}{k}}$ is the $k$-th root of unity. This character table has complex valued elements so we use (4-13) to obtain $|\alpha_m(t)|_{opt}$, so

$$|\alpha_0(t)|_{opt} = |\alpha_k(t)|_{opt} = \frac{1}{2k}\sum_{l=1}^{2k} 1 \times 1 = 1$$

$$|\alpha_m(t)|_{opt} = \frac{1}{2k}\left(1+1+\sum_{l=1}^{k-1} 1 \times \left|\omega^{ml} + \bar{\omega}^{ml}\right|\right) = \frac{1}{k}\left(1+\sum_{l=1}^{k-1}\left|Cos(\frac{ml\pi}{k})\right|\right)$$

Using (3-16) we get

$$C^{opt}_{ml,ml'} = \frac{2}{k^2}\left(1+\sum_{l=1}^{k-1}\left|Cos(\frac{ml\pi}{k})\right|\right)^2$$

Regarding (4-15) we clearly observe that $\{J_l\}^m_{opt} = \{J_l\}^{\bar{m}}_{opt}$.

4.2.3 $SL(2,3)$

The group $SL(2,p)$ where p is a prime number is generated by a generator $\begin{pmatrix} a & b \\ c & d \end{pmatrix}$ with the following relations [21]

$$SL(2,p) = \left\{ \begin{pmatrix} a & b \\ c & d \end{pmatrix} : a,b,c,d \in \mathbb{Z}_p, ad - bc = 1 \right\}$$

For the sake of simplicity we investigate optimal state transfer over UNGAS of $SL(2,3)$. The number of group elements of $SL(2,3)$ is $n = 24$ and the character table of this group is as follows:

| $g_i$ | $e$ | $g_1$ | $g_2$ | $g_3$ | $g_4$ | $g_5$ | $g_6$ |
|---|---|---|---|---|---|---|---|
| $\kappa_i$ | 1 | 1 | 6 | 4 | 4 | 4 | 4 |
| $\chi_1$ | 1 | 1 | 1 | 1 | 1 | 1 | 1 |
| $\chi_2$ | 1 | 1 | 1 | $\omega$ | $\omega^2$ | $\omega^2$ | $\omega$ |
| $\chi_3$ | 1 | 1 | 1 | $\omega^2$ | $\omega$ | $\omega$ | $\omega^2$ |
| $\chi_4$ | 3 | 3 | $-1$ | 0 | 0 | 0 | 0 |
| $\chi_5$ | 2 | $-2$ | 0 | $-\omega^2$ | $-\omega$ | $\omega$ | $\omega^2$ |
| $\chi_6$ | 2 | $-2$ | 0 | $-\omega$ | $-\omega^2$ | $\omega^2$ | $\omega$ |
| $\chi_7$ | 2 | $-2$ | 0 | $-1$ | $-1$ | 1 | 1 |

where $\omega := e^{\frac{2\pi i}{3}}$. This character table has complex valued elements so we use (4-13) to obtain $|\alpha_m(t)|_{opt}$,

$$|\alpha_0(t)|_{opt} = |\alpha_1(t)|_{opt} = \frac{1}{24}(3 \times 1 + 3 \times 3 + 3 \times 2 \times 2) = 1$$

$$|\alpha_2(t)|_{opt} = \frac{1}{24}(3 \times 1 + 3 \times 1) = \frac{1}{4}$$

$$|\alpha_3(t)|_{opt} = |\alpha_4(t)|_{opt} = \frac{1}{24}\left(1 \times 1 + 1 \times |\omega + \omega^2| + 0 + +2 \times |-\omega - \omega^2| + 2 \times 1\right) = \frac{1}{4}$$

$$|\alpha_5(t)|_{opt} = |\alpha_6(t)|_{opt} = \frac{1}{24}\left(1 \times 1 + 1 \times |\omega + \omega^2| + 0 + 2 \times |\omega + \omega^2| + 2 \times 1\right) = \frac{1}{4}$$

As pointed out in (4-16)
these results show that the conjugate strata have equal optimal excitation probability. Using (3-16) we get

$$C^{opt}_{2l,2l'} = C^{opt}_{3l,3l'} = C^{opt}_{4l,4l'} = C^{opt}_{5l,5l'} = C^{opt}_{6l,6l'} = \frac{1}{8}$$

### 4.2.4 $V_{8k}$

The group $V_{8k}$, where $k$ is an odd integer number [21], is generated by two generators $a$ and $b$ with the following relations:

$$V_{8k} = \{a,b : a^{2k} = b^4 = 1, ba = a^{-1}b^{-1}, b^{-1}a = a^{-1}b\}$$

The number of group elements of $V_{8k}$ is $n = 8k$ and the character table of this group is as follows:

| $g_i$ | $e$ | $b^2$ | $a^{2r+1}$ $(0 \leq r \leq k-1)$ | $a^{2s}$ $(1 \leq s \leq k - 1/2)$ | $a^{2s}b^2$ | $b$ | $ab$ |
|---|---|---|---|---|---|---|---|
| $\kappa_i = |C_i|$ | 1 | 1 | 2 | 2 | 2 | $2k$ | $2k$ |
| $\chi_1$ | 1 | 1 | 1 | 1 | 1 | 1 | 1 |
| $\chi_2$ | 1 | 1 | 1 | 1 | 1 | $-1$ | $-1$ |
| $\chi_3$ | 1 | 1 | $-1$ | 1 | 1 | 1 | $-1$ |
| $\chi_4$ | 1 | 1 | $-1$ | 1 | 1 | $-1$ | 1 |
| $\psi_j$ $(0 \leq j \leq k-1)$ | 2 | $-2$ | $\omega^{2j(2r+1)}$ $-\omega^{-2j(2r+1)}$ | $\omega^{4js}$ $+\omega^{-4js}$ | $-\omega^{4js}$ $-\omega^{-4js}$ | 0 | 0 |
| $\phi_j$ $(1 \leq j \leq k-1)$ | 2 | 2 | $\omega^{2j(2r+1)}$ $+\omega^{-2j(2r+1)}$ | $\omega^{2js}$ $+\omega^{-2js}$ | $\omega^{2js}$ $+\omega^{-2js}$ | 0 | 0 |

where $\omega := e^{\frac{\pi i}{k}}$. This character table has complex valued elements as well so we use (4-13) to obtain $|\alpha_m(t)|_{opt}$ over different strata. Note that the complex valued rows correspond to $\psi_j : 1 \leq j \leq k-1$, but $\psi_0$ is real. So the results will be

$$\left|\alpha_{a^{2r+1}}(t)\right|_{opt} = \frac{1}{8k}\left(4 \times 1 \times 1 + 0 + 2 \times \sum_{j=1}^{\frac{k-1}{2}}\left|\text{Re}al\left(2i\text{Sin}\left(\frac{2j\pi(2r+1)}{k}\right)\right)\right| + 2 \times \sum_{j=1}^{k-1}\left|\text{Cos}\left(\frac{2j\pi(2r+1)}{k}\right)\right|\right) =$$

$$\frac{1}{2k} + \frac{1}{2k}\sum_{j=1}^{\frac{k-1}{2}}\left|\text{Cos}\left(\frac{2j\pi(2r+1)}{k}\right)\right|$$

$$\left|\alpha_{a^{2s}}(t)\right|_{opt} = \left|\alpha_{a^{2s}}(t)\right|_{opt} = \frac{1}{8k}\left(4 \times 1 \times 1 + 2 \times 2 + 2 \times \sum_{j=1}^{\frac{k-1}{2}}2\left|\text{Cos}(4js)\right| + 2 \times \sum_{j=1}^{k-1}\left|\text{Cos}(2js)\right|\right) =$$

$$\frac{1}{k} + \frac{1}{2k}\sum_{j=1}^{\frac{k-1}{2}}\left(\left|\text{Cos}(4js)\right| + \left|\text{Cos}(2js)\right|\right)$$

$$|\alpha_b(t)|_{opt} = |\alpha_{ab}(t)|_{opt} = \frac{1}{8k}(4 \times 1 \times 1 + 0) = \frac{1}{2k}$$

$$C^{opt}_{a^{2s}l, a^{2s}l'} = \frac{1}{2k^2}\left(1 + \sum_{j=1}^{\frac{k-1}{2}}\left|Cos\left(\frac{2j\pi(2r+1)}{k}\right)\right|\right)^2$$

$$C^{opt}_{a^{2s}l, a^{2s}l'} = \frac{1}{2k^2}\left(2 + \sum_{j=1}^{\frac{k-1}{2}}(|Cos(4js)| + |Cos(2js)|)\right)^2$$

$$C^{opt}_{bl,bl'} = C^{opt}_{abl,abl'} = \frac{1}{2k^2}$$

In this section we obtained analytical formulas for creation of optimal entanglement between vertex pairs locating in the same strata and practice them in some examples. In the next section we study the relations governing over entanglement of vertex pairs locating in different strata and compare them with the above relations.

## 5  Optimal Entanglement Generation between Different Strata

We have given analytical expressions for optimal entanglement between vertices located on the same stratum. However such analytical results do not hold when the vertices are located on different strata we need a different approach is offered in the present section. At first step some hidden nonlinear constraints among the set of $\{\alpha_m(t)\}$ are clarified. Existence of these constraints means that the optimization problem should be treated as a "nonlinear constraint optimization problem". Thus we use some suitable numerical method in this context. Moreover two different upper bounds are presented. And at the end of this section we shall analyze the dihedral group -as an example- and the strength of presented upper bounds is checked by comparing them to the numerical solution results.

Multiplying $\alpha_m(t)$ in (3-13) by $P_{km}$ and sum over $m$, we find:

$$\sum_{m=0}^{d} P_{km}\alpha_m(t) = e^{i\theta_k} \tag{5-1}$$

Since throughout the paper the systems with real valued $\theta_k$'s have been under consideration the norm of (5-1) equals unity. Therefore the following quadratic constraints on the set of $\{\alpha_k(t)\}$ is obtained

$$\sum_{m=0}^{d}\sum_{m'=0}^{d} P_{km}\overline{P_{km'}}\alpha_m(t)\overline{\alpha_{m'}(t)} = 1 \tag{5-2}$$

Let us now consider the optimization of target function $T(t)$ (3-15) under the nonlinear constraints (5-2). Due to the quadratic form of obtained constraints the sequential quadratic programming method (SQP) is best suited for numerical solution of optimal values, because SQP gives the best results when the nonlinear constraints are of this type [26]. But first we derive some bounds for the said function.

## 5.1 Some upper bounds over optimal entanglement of vertices of different strata

The first upper bound results immediately from the optimal entanglement of vertices of the same strata which calculated analytically in previous section. In fact we have

$$C_{il,jk}^{opt} = 2|\alpha_i(t)\alpha_j(t)|_{opt} = 2|\alpha_i(t_{ij}^{opt})\alpha_j(t_{ij}^{opt})| =$$
$$2|\alpha_i(t_{ij}^{opt})||\alpha_j(t_{ij}^{opt})| \leq 2|\alpha_i(t_i^{opt})||\alpha_j(t_j^{opt})_{opt}| \quad (5\text{-}3)$$

Where $t_i^{opt}, t_j^{opt}$, and $t_{ij}^{opt}$ are the times of optimization of $|\alpha_i(t)|_{opt}$, $|\alpha_j(t)|_{opt}$, and $|\alpha_i(t)\alpha_j(t)|_{opt}$ respectively. For $\kappa_i, \kappa_j \neq 1$, the inequality (5-3) takes the form

$$C_{ij\_\max} \leq \sqrt{C_{ii\_\max} C_{jj\_\max}} \quad (5\text{-}4)$$

Where $C_{ij\_\max}$ depicts optimal entanglement between vertices over $i$ and $j$ strata. This inequality shows that the greatest amount of entanglement could be achieved between vertex pairs located in the same stratum.

The second upper bound is attained by considering the probability conservation rule (3-5). Assume the ultimate situation where excitations are completely localized in the targeted strata $i$ and $j$. Regarding (3-14) we use Lagrange multipliers to obtain an upper bound. The Lagrange function is

$$\Lambda(|\alpha_i|,|\alpha_j|,\lambda) = 2|\alpha_i\alpha_j| - \lambda(\kappa_i|\alpha_i|^2 + \kappa_j|\alpha_j|^2 - 1) \quad (5\text{-}5)$$

So the optimal value of the target function is obtained as

$$C_{ij\_UB} = \frac{1}{\sqrt{\kappa_i \kappa_j}} \quad (5\text{-}6)$$

According to (5-6) to achieve rather larger amount of entanglement we must consider groups with strata that contains a few vertices. For example cyclic groups discussed at 4.2.2 seem well suited because $\kappa_i \leq 2$. In the next subsection we check the strength of two upper bounds (5-4) and (5-6) by comparing with numerical results.

## 5.2 Optimal entanglement over UNGAS of $D_6$ group

As an example, in this subsection we investigate the magnitudes of optimal entanglement between different strata of the Dihedral group $D_6$ by applying the results of section 5 and subsection 5.1. So here we present the results of SQP numerical solution on the magnitude of optimal entanglement between different strata. Moreover we compute the upper bounds over optimal entanglement between different strata and compare them with the numeric results in order to check the strength of these upper bounds.

In order to calculate upper bounds (5-4) and (5-6), $|\alpha_i|_{opt}$ must be identified which is calculated in subsection 4.2.1. We present the numeric values and the upper bounds in the following table

|                   | $C_{01}$ | $C_{02}$ | $C_{12}$ |
|-------------------|----------|----------|----------|
| first upper bound | 1.3333   | 0.6667   | 0.4444   |
| second upper bound| 0.7071   | 0.5774   | 0.4082   |
| numerical results | 0.7071   | 0.4873   | 0.2886   |

The results show, in this example, the first upper bound is a weak estimation of optimal entanglement between different strata especially in estimation of $C_{01}$. The second upper bound is more stringent than the first one. In particular this upper bound gives exact result of optimal entanglement between the zero and first strata. Since in obtaining the second upper bound we use the assumption that the whole of excitations are gathered in the targeted strata, we deduce that the dynamics of the system permits this total localization to happen.

As we calculated in 4.2.1 the optimal entanglement among vertices located in the same strata is $C_{11\_max} = \frac{8}{9}$ and $C_{22\_max} = \frac{2}{9}$. The larger amount of $C_{11\_max}$ than other entanglement optimal values is in accordance with inequality (5-4).

# 6  Summary
We have discussed optimal entanglement provision over symmetric networks introduced in this paper as UNGAS. It is noteworthy that recently UNGAS has received attention due to their capability in perfect state transfer [10]. Initially entanglement between vertex pairs belonging to the same stratum is calculated for general couplings and initial states. The first result obtained is that the optimal initial state is a simple local excited state that can be prepared simply by applying a local magnetic field at controlled site in the cooling process. Moreover we obtained an analytic formula for optimal entanglement achievable between vertex pairs located in the same stratum. For those pairs located in different strata two upper bounds are presented. These upper bounds indicate the amount of entanglement achieved between pairs located in different strata is lower than vertex pairs over the same stratum. We also find that achieving larger amount of entanglement between different vertices the strata must contain a few elements. Finally the efficacy of these upper bounds is checked numerically in an example. Systems considered here may be realized by arrays of quantum dots [22] or atoms trapped in optical lattices [23]. This is a highly challenging problem related to implementation of nanoscale quantum information processors. This study clarifies in what extents and which circumstances UNGAS may be used in future quantum information processors for providing entanglement between different nodes.